# Optimization of an electret-based energy harvester

S Boisseau[1], G Despesse[1] and A Sylvestre[2]

[1] CEA/LETI, 17 avenue des martyrs, Minatec, Grenoble, France
[2] G2ELab, CNRS, 25 avenue des Martyrs, Grenoble, France

**Abstract.** Thanks to miniaturisation, it is today possible to imagine self-powered systems that use vibrations to produce their own electrical energy. Many energy-harvesting systems already exist. Some of them are based on the use of electrets: electrically charged dielectrics that can keep charges for years. This paper presents an optimisation of an existing system and proves that electret-based electrostatic energy scavengers can be excellent solutions to power microsystems even with low-level ambient vibrations. Thereby, it is possible to harvest up to 200µW with vibrations lower than 1G of acceleration (typically 50µm$_{pp}$ at 50Hz) using thin SiO$_2$ electrets with an active surface of 1 cm² and a mobile mass of 1g. This paper optimises such a system (geometric, electrostatic and mechanical parameters), using FEM (Finite Element Method) software (Comsol Multiphysics) and Matlab to compute the parameters and proves the importance of such an optimisation to build efficient systems. Finally, it shows that the use of electrets with high surface potential is not always the best way to maximise output power.

## 1. Introduction

Today, one of the goals of Micro-Electro-Mechanical Systems (MEMS) researchers is to build an autonomous microsystem made of sensors, actuators, data processing units and an energy source to power all these elements. Until now, this energy is provided by batteries whose main disadvantage is the lifetime (refilling). Thanks to miniaturisation, systems are consuming less and less energy giving them the opportunity to produce their own electricity by extracting their surrounding ambient environment energy. Presently, existing energy scavenging systems for microsystems are focused on three types of energy: (i) mechanical energy harvesting using piezoelectric, electrostatic, or electromagnetic transducers (a good review can be found in [1]); (ii) thermal energy harvesting using Seebeck's effect [2] or (iii) solar energy harvesting using photovoltaic cells [3]. In this paper, we focus on vibration energy harvesting. All conversion principles (piezoelectric, electromagnetic or electrostatic) have both pros and cons and the choice of the technology to use to harvest energy depends on the application. For example, for an application in ambient energy scavenging, electrostatics systems seem to be particularly advantageous especially due to their ability to work under low frequencies and/or on wide frequency bandwidths. They also take advantage of miniaturisation. Indeed, electrostatic systems work thanks to the displacement of an electrode compared to another one, generally separated by an air gap. The power output is linked to its thickness: decreasing the air gap thickness thanks to miniaturisation increases the power output. Electret-based energy harvesters are part of electrostatic generators. Electrets are mainly known for their use in microphones but their potential in energy harvesters has already been proven [4, 5]. Demonstrators using this technology have already been developed [6–7], and, recently, prototypes from Omron Company [8] and Sanyo [9] have been built and are currently tested on real environments

(roads, elevators...).

In order to optimise systems and to increase energy harvesting, it is necessary to develop models. Generally, the electrostatic system is modelled by the simple laws of electrostatic such as the plane capacitor [5, 10]. Nevertheless, some researchers have worked on models for other electrostatic systems such as comb-drive scavengers (with or without electrets). For example, Sterken *et al.* [11] have used an equivalent electrical circuit to represent both the electrostatic and the mechanical part of their energy harvester. By using this circuit, they have proven that it was also possible to optimise the structure. Contrary to Sterken *et al.*, Peano *et al.* [12] have used the standard laws of electricity and mechanics to make a model of their system. They have also made a procedure to optimise their energy harvester. Another model from Tvedt et al. [13], investigates the effects of linear and non-linear models of an electrostatic in-plane overlap varying energy harvester using PSPICE: like *Peano et al.* the mechanical system is solved by transforming it into its equivalent electrical circuit. Generally, the main problem of all these models is to neglect the fringe effects that appear and become predominant when working with small dimensions. Nevertheless, Le *et al.* [14] have developed an analytical model for computing the capacitance variation for an in-plane overlap plate transducer. Based on an analytic expression, it partially considers fringe effects and leads to a better modelling of the structure.

In this study, we have decided to use a Finite Element Method (FEM) software (Comsol Multiphysics) to compute the capacitance of the system considering all fringe effects. Therefore, it is possible to obtain a better result on the capacitance estimation. Then, by parameterizing the geometry of the structure, an optimisation using both Comsol Multiphysics and Matlab has been implemented and has proven the importance of the geometric parameters to maximise the power output. A method is then developed to design efficient energy scavengers with electrets. This paper firstly introduces the existing systems and the energy harvester studied in the next sections. This latter can be divided into two parts: an electrostatic converter and a mechanical structure. In section 3, we present our model of the electrostatic converter using FEM, that will be optimised in section 4. Section 5 is aimed at getting the optimisation of the entire structure and the harvestable power with ambient vibrations.

## 2. Introducing the structure and the parameters

### 2.1 State of the art

Since 1978, and the first electret generator made by Jefimenko [4], many electrostatic energy harvesters using electrets have been built. Table 1 gives some examples of those electret generators:

**Table 1.** Electrets energy harvesters.

| Author | Ref | Vibrations / Rotations | Active Surface | Electret Potential | Output Power | Figure of merit $\chi$ |
|---|---|---|---|---|---|---|
| Jefimenko | [4] | 6000 rpm | 730 cm² | 500V | 25 mW | 5.94E-08 |
| Tada | [15] | 5000 rpm | 90 cm² | 363V | 1.02 mW | 2.76E-07 |
| Boland | [5] | 4170 rpm | 0.8 cm² | 150V | 25 µW | 1.47E-04 |
| Genda | [7] | 1'000'000 rpm | 1.13 cm² | 200V | 30.4 W | 6.51E-06 |
| Boland | [10] | 1mm$_{pp}$@60Hz | 0.12 cm² | 850V | 6 µW | 3.73E-02 |
| Tsutsumino | [6] | 2mm$_{pp}$@20Hz | 4 cm² | 1100V | 38 µW | 4.79E-02 |
| Lo | [16] | 2mm$_{pp}$@60Hz | 4.84 cm² | 300V | 2.26 µW | 8.72E-05 |
| Sterken | [17] | 2µm$_{pp}$@500Hz | 0.09 cm² | 10V | 2nW | 7.17E-03 |
| Lo | [18] | 1mm$_{pp}$@50Hz | 6 cm² | 1500V | 17.98 µW | 3.87E-03 |
| Omron | [8] | 1.2mm$_{pp}$@20Hz | 4 cm² | 700V | 10 µW | 3.50E-02 |
| Zhang | [19] | 2mm$_{pp}$@9Hz | 4 cm² | 100V | 0.13 pW | 1.80E-09 |
| Yang | [20] | 5µm$_{pp}$@560Hz | 0.3 cm² | 400V | 46.14 pW | 5.65E-06 |
| Suzuki | [21] | 2mm$_{pp}$@37Hz | 2.33 cm² | 450V | 0.28 µW | 9.56E-05 |
| Sakane | [22] | 1.2mm$_{pp}$@20Hz | 4 cm² | 640V | 0.7 mW | 2.45E+00 |
| Sanyo | [9] | 50mm$_{pp}$@2Hz | 9 cm² | | 40µW | 3.58E-02 |
| Halvorsen | [23] | 5.6µm$_{pp}$@596Hz | 0.48 cm² | | 1µW | 5.06E-02 |
| Kloub | [24] | 0.16µm$_{pp}$@1740Hz | 0.42 cm² | 25V | 5µW | 1.42E+01 |
| Edamoto | [25] | 1mm$_{pp}$@21Hz | 3 cm² | 600 V | 12µW | 6.97E-02 |
| Miki | [26] | 0.2mm$_{pp}$@63Hz | 3 cm² | 180V | 1µW | 5.37E-03 |
| Honzumi | [27] | 18.7µm$_{pp}$@500Hz | 0.01 cm² | 52V | 90 pW | 3.32E-05 |

It is hard to find a criterion to compare systems since they do not work with the same vibrations, the same electrets and above all the same mass. A good way to normalise the output powers is to build a figure of merit. We have chosen to build our figure of merit $\chi$ by dividing the output power ($P$) by the active surface ($S$) and the available mechanical power from the environment. This latter is got by the product of the inertial force ($Y\omega^2$) and the speed of the moving part of the energy harvester ($Y\omega$)[1], where $Y$ is the amplitude of the vibrations and $\omega$ the angular frequency of these vibrations.

$$\chi = \frac{P}{SY^2\omega^3} \tag{1}$$

The architecture studied by Sakane *et al.* [22] is the second one with our figure of merit but it works with ambient vibrations and, consequently, has been chosen for the rest of this study. Nevertheless, the method that will be developed in the next parts could be adapted to other systems by changing only the computation of the capacitance in the FEM software.

*2.2 Presentation of the energy harvester*
- Mechanical model of vibration energy harvesters

All resonant structures that harvest energy from vibrations can be modelled as a moving mass ($m$) maintained in a fixed frame by a spring ($k$) and amortised by forces. The vibrations of the environment $y(t)$ (amplitude: $Y$, frequency: $f$) induce a displacement $x(t)$ (amplitude: $X$, frequency: $f$) of the moving mass ($m$) relative to the frame. Part of the kinetic energy of the moving mass is lost due to mechanical damping ($f_{mec}$) modelled as a viscous friction force $f_{mec} = b_m\dot{x}$, while the other part is converted into electricity, which is modelled by an electrostatic force ($f_{elec}$), in electrostatic energy harvesters (figure 1). As ambient vibrations are generally low amplitude, this mass-spring structure enables to take advantage of a phenomenon of resonance that can amplify the amplitude of vibrations perceived by the mobile mass. According to the fundamental principle of dynamics (gravity is neglected):

$$m\ddot{x} + b_m\dot{x} + kx + f_{elec} = -m\ddot{y} \tag{2}$$

---
[1] For rotating energy scavengers, we have taken Y=R, where R is the radius of the energy scavenger.

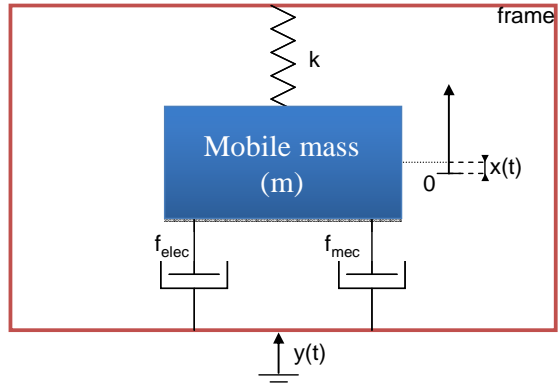

**Figure 1.** Mechanical system

- Conversion using electrets

As the mechanical part is described, we can now study the principles of the conversion. In the system introduced in figure 2, the electret has a fixed charge $Q_i$ and is deposited on an electrode. A counter-electrode is placed opposite the electrode and spaced with an air gap. Because of electrostatic induction and conservation of charges, $Q_i = Q_1 + Q_2$ at any moment, where $Q_1$ is the charge on the electrode and $Q_2$ the charge on the counter-electrode.

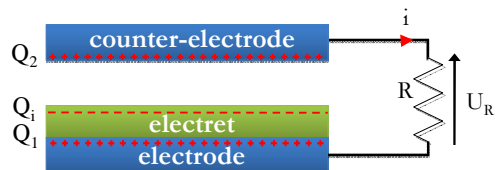

**Figure 2.** Electrostatic converter using electret

Vibrations from the environment induce changes in the geometry of the capacitor (e.g. counter-electrode moves parallel to the electrode) and the value of the capacitance changes. The charges on the electrode and on the counter-electrode reorganise themselves through the load $R$. This generates a charge variation and a current circulation through the load: mechanical energy is turned into electricity.

- Complete electromechanical system

To build an electrostatic energy harvester, the electrostatic converter (figure 2) must be integrated into the mechanical system (figure 1). Figure 3 gives a schematic structure of the complete electromechanical system (studied in particular by the University of Tokyo [22].)

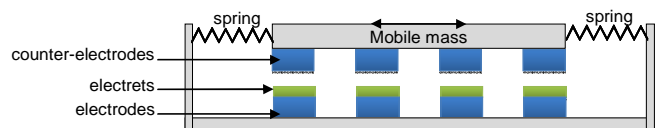

**Figure 3.** Complete energy harvester

The system is composed of two plates in front of each other. The upper plate moves respectively to the lower one thanks to springs when a vibration occurs. The relative displacement between the two plates is used to convert mechanical energy into electricity thanks to a converter made of patterned electrodes, counter-electrodes and electrets.

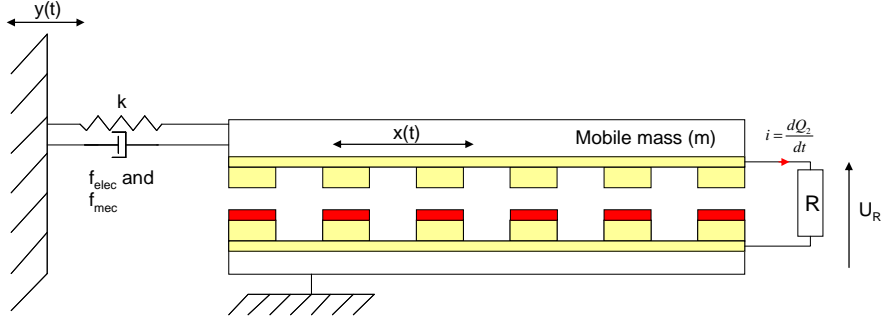
**Figure 4.** Parameters of the complete system.

The system is parameterised in figure 4: $m$ is the mobile mass, $k$ the spring stiffness, $x(t)$ the displacement of the upper electrodes relative to the lower ones, $y(t)$ the ambient vibrations, $f_{mec}$ the friction forces and $f_{elec}$ the electrostatic forces which represents the loss of kinetic energy converted into electricity. To maximise the output power, the electrostatic forces must be optimised: if they are too weak, too little energy is converted from ambient vibrations. If they are too strong, too much energy is converted per displacement unit and finally there is not enough kinetic energy to maintain the displacements of the mobile mass, which is eventually blocked: there is little relative displacement between the upper and the lower electrodes of the converters, and little energy is converted. To maximise the output power, both the electrostatic converter and the mechanical part must be optimised.

## 3. Modelling of the electrostatic converter

Before optimising the electrostatic converter of the energy harvester introduced in figure 3, it is necessary to develop an accurate model of its variable capacitor.

### 3.1 Present modelling – Boland's formulas – Defects

Boland [5] already did a modelling of this kind of electrostatic converter in 2003. This simple analytic model consists in a variable capacitor in which the upper electrode moves parallel to the lower electrode (figure 5).

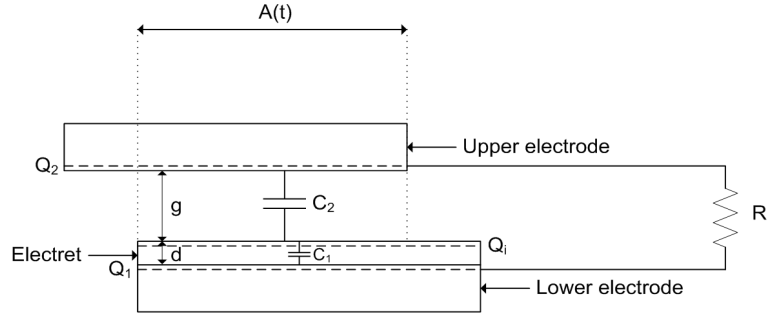
**Figure 5.** Parameters of the simple analytic model.

$A(t)$ is the surface of coincidence of the two electrodes, $d$ and $g$ are respectively the thickness of the electret and the thickness of the gap between the electret and the upper electrode. The results of Boland in terms of converted power and optimal load are summarised in (3):

$$P_{max} = \frac{\sigma^2 \frac{d}{dt}(A(t))}{4\varepsilon_0 \varepsilon_1 \left(\frac{\varepsilon_1 g}{\varepsilon_2 d} + 1\right)} \quad \text{and} \quad R_{max} = \frac{1}{\varepsilon_0 \frac{d}{dt}(A(t))}\left(\frac{d}{\varepsilon_1} + \frac{g}{\varepsilon_2}\right) \tag{3}$$

with $\varepsilon_0$, $\varepsilon_1$ and $\varepsilon_2$, respectively the permittivity of vacuum, the dielectric constant of the electret and the dielectric constant of the dielectric which separates the upper electrode from the electret. $P_{max}$ is the maximum power output, $R_{max}$ the optimal load, and $\sigma$ the surface charge density of the electret. These results are obtained by using the standard laws of electricity (Kirchhoff's laws) and does not consider fringe effects: they are only valid when the surface of the capacitor is large compared to the thickness of the air gap ($A(t) > 20 \times g$).

Therefore, the major defect of this analytic model is to neglect fringe effects. This leads to an overestimation of the capacitance variation between the two electrodes. Moreover, as the converted power is intimately linked to the maximum capacitance variation ratio ($\Delta C_{max} = C_{max}/C_{min}$) of the system when the upper electrode moves, by neglecting fringe effects, the output power is overestimated. The use of FEM allows consideration of fringe effects and better approximation of the output power.

### 3.2 FEM model

The goal of FEM simulation is to compute the capacitance of the electrostatic converter. This system is simulated by using the electrostatic module (*es module*) of Comsol Multiphysics and the capacitance is deduced from the electrical energy ($W_e$) given by the software when the capacitance (*C*) is charged under $V_c$:

$$W_e = \frac{1}{2} C V_c^2 \tag{4}$$

Figure 6(a) shows the boundary conditions and figure 6(b), the electric field given by Comsol Multiphysics. The potential of the upper electrodes is set to $+V_c$, the potential of the lower electrodes is set to *GND*. The boundary conditions of electrets are set to *continuity*.

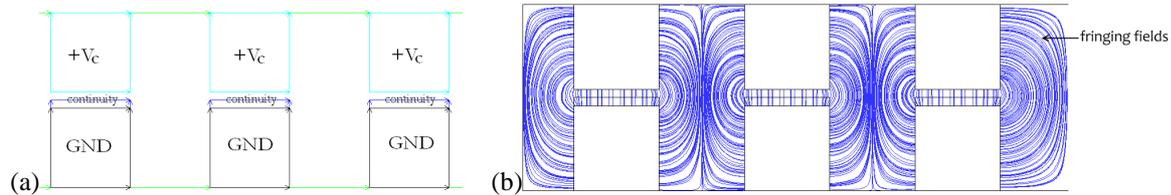

**Figure 6.** (a) FEM model. (b) Solution: electric field

Figure 6(b) proves the existence of fringe effects on the sides of the bumps: part of the electric field is outside the gap.

### 3.3 Comparison between the analytic model and the FEM model - Capacitance

The chart in figure 7 shows the comparison between the theoretical formula (5) and the simulation results under Comsol Multiphysics, which takes into account fringe effects. It presents an example of the capacitance value as a function of the relative displacement between the upper and the lower electrodes *x(t)*.

$$C = \varepsilon_0 A(t) \frac{1}{g + \frac{d}{\varepsilon}} \tag{5}$$

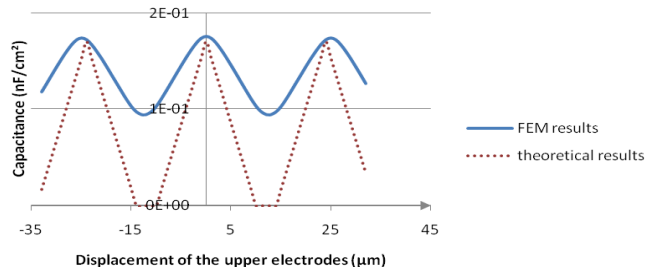

**Figure 7.** Comparison of the capacitance obtained by FEM and theory.

While $\Delta C_{max}$ is huge (varies from $C_{max}$ *[A(t) is maximal]* to nearly 0 *[A(t) equals 0]*) when the system is modelled by the simple analytic model, it appears that, in fact, it is difficult to reach a capacitance variation higher than 3. Figure 7 shows that the capacitance variation is significantly overestimated when using the theoretical formulas and confirms that fringing fields cannot be ignored.

## 3.4 Capacitance as a sinusoidal function

The biggest problem with FEM models is the time needed for the simulation. In fact, it would take too much time to compute the capacitance variation as a function of the displacement point by point. To limit the simulation duration, the problem has to be reduced to the determination of some values and the use of an interpolation function. The FEM results curve presented in figure 7 looks like a cosine function with two extrema: $C_{min}$ and $C_{max}$. Therefore, with only this function of interpolation and $C_{min}$ and $C_{max}$, one can estimate the capacitance value as a function of the displacement. Obviously, $C_{min}$ and $C_{max}$ depend on the geometrical parameters of the structure. We have chosen to call them: $g$, $d$, $b$, and $e$, where $g$ is the air gap between the upper and the lower part of the structure, $d$ the thickness of the electret, $b$ the width of the bumps and $e$ the width of the spaces between them (figure 8). As for the height of the bumps, it has small effect compared with the other parameters as soon as they are at least ten times bigger than the gap.

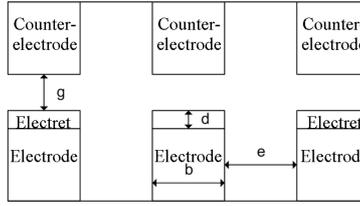

**Figure 8.** Geometrical parameters of the structure.

Consequently, $C(x)$ can be expressed as:

$$C(x) = \frac{C_{max} + C_{min}}{2} + \left(\frac{C_{max} - C_{min}}{2}\right) \times \cos\left(\frac{2\pi x}{e+b}\right) \quad (6)$$

To validate this expression (6), different configurations have been tested and compared to point-by-point FEM results.

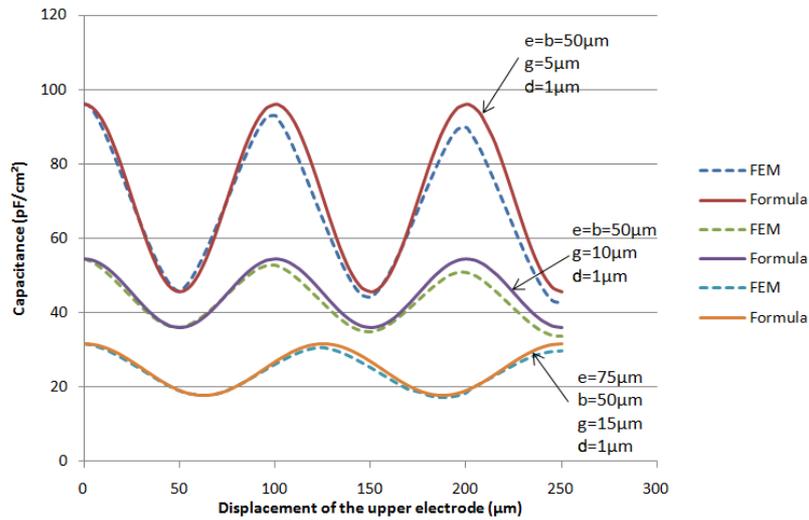

**Figure 9.** Comparison of the capacitance obtained by the formula and FEM software.

Figure 9 shows the good agreement between the simulation (point-by-point) and calculation (from $C_{min}$ and $C_{max}$) results. The limited number of bumps (30 bumps) in this simulation can explain the differences that appear for big displacements, while the use of the function supposes an infinite number of bumps. Thus, the capacitance of the system can be well fitted by an analytic function in which only two parameters have to be computed with the FEM software. This has two advantages: first, it is fast and more correct than the theoretical formula to estimate the output power and secondly, it is easier to implement under a numerical solver. Thanks to this formula, we can optimise the electrostatic converter.

## 4. Optimisation of the electrostatic converter

The goal of this optimisation is to maximise the converted power of the electrostatic converter for a given surface and for a given relative displacement of the upper electrodes compared to the lower electrodes ($x(t)$). The whole structure working with ambient vibrations will be studied in section 5.

### 4.1 Ordinary Differential Equation (ODE)

The output power of the electrostatic converter is obtained from the current that circulates through the load ($R$). Boland [5] has proven that the ODE of the system is:

$$\frac{dQ_2}{dt} = \frac{\sigma d}{R\varepsilon\varepsilon_0} - \frac{Q_2}{A(t)R}\times\left(\frac{d}{\varepsilon\varepsilon_0}+\frac{g}{\varepsilon_0}\right) = \frac{V}{R} - \frac{Q_2}{A(t)R}\times\left(\frac{d}{\varepsilon\varepsilon_0}+\frac{g}{\varepsilon_0}\right) \quad (7)$$

where $Q_2$ is the charge on the upper electrode, $\sigma$ the surface charge density of the electret, $A(t)$ the surface of coincidence between the upper and the lower electrodes, $d$ the thickness of the electret, $g$ the thickness of the air gap, $\varepsilon$ the relative permittivity of the electret, $\varepsilon_0$ the permittivity of vacuum and $V$ the surface voltage of the electret. To add fringe effects, the capacitance $\left(\frac{d}{\varepsilon\varepsilon_0}+\frac{g}{\varepsilon_0}\right)\times\frac{1}{A(t)}$ is replaced by its equivalent computed by FEM:

$$\frac{dQ_2}{dt} = \frac{\sigma d}{R\varepsilon\varepsilon_0} - \frac{Q_2}{R}\times\left[\frac{1}{C}\right]_{FEM} = \frac{V}{R} - \frac{Q_2}{R}\times\left[\frac{1}{C}\right]_{FEM} \quad (8)$$

Equation (8) is implemented in Matlab/Simulink and the average output power is calculated with (9), where $t_{end}$ is the end time of the simulation:

$$P_{average} = \frac{1}{t_{end}}\int_0^{t_{end}} R\left(\frac{dQ_2}{dt}\right)^2 \quad (9)$$

### 4.2 Validation on experimental data coming from Tsutsumino [6]

Our model and (8) were tested on experimental data coming from Tsutsumino [6] ($f$=20Hz, $X$=1mm, $g$=100µm, $d$=15µm, $e$=1mm, $b$=1mm, $V$=-950V, $R$=60MΩ). The simulation (figure 10) and the experimental data give the same magnitudes of power: the output power expected by the simulation is 46µW while it is 37.7µW in [6] and the shape of the curves are more or less the same [6].

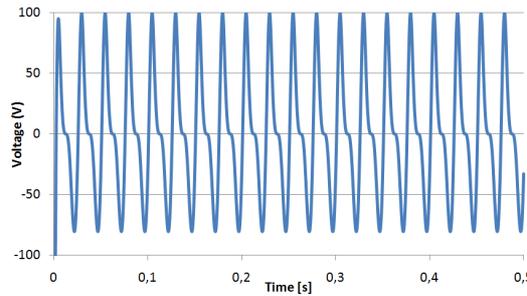

**Figure 10.** Output voltage obtained by our model

This confirms that the results obtained with our model are similar to experimental data. The small differences observed are probably due to uncertainties or differences between the values given in the paper and the real values of the experiment.

### 4.3 Output power as a function of X, f, V and S

The problem is now to get a simple expression of the converted power as a function of the relative displacement of the upper electrodes ($x(t)$). Equation (8) was numerically solved in Matlab/Simulink. First, the maximum of converted power for different values of the surface voltage of the electret $V$ has been computed using (8). Each time, the optimal load is found using an optimisation program. The other parameters ($X, f, e, b, g, d$) are kept constant. The results are presented in figure 11.

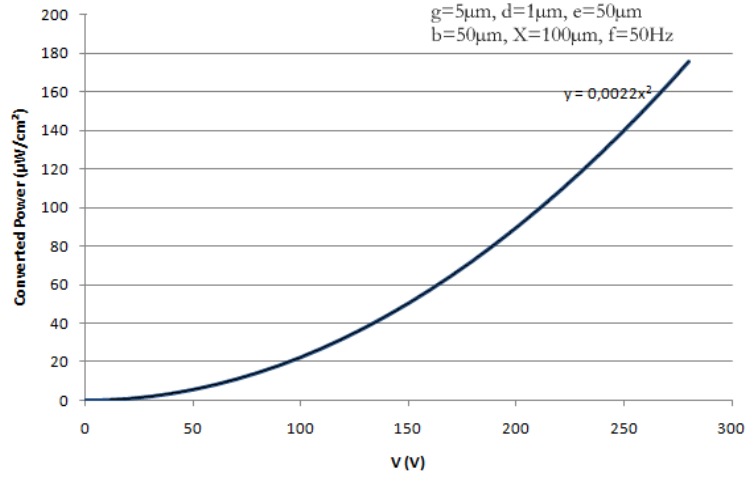

**Figure 11.** Converted power as a function of *V*, the surface voltage of the electret.

Boland's formula says that the output power is a function of $V^2$: this is what has been found with our simulation. Moreover, other simulations with other variables have proven that, for a given displacement of the upper electrodes, the maximum converted power $P_{out}$ (output power when the load is optimised) can be easily expressed as a function of *X*, the relative displacement amplitude of the upper electrodes compare to the lower electrodes, *f* its frequency, *V* the surface voltage of the electret and *S* the surface of the active part (bumps+spaces).

$$\frac{P_{out}}{X\,2\pi f V^2 S} = \text{constant} = P_N \qquad (10)$$

As $P_N$ is a constant depending only on the geometrical parameters, it can be considered as a figure of merit able to compare the efficiency of the capacitance geometries. It will be referred as the Normalised Output Power.

*4.4 A first limit on V coming from Paschen's effect*

To maximise the converted power, it seems that *V* has to be taken as high as possible. Nevertheless, by decreasing the air gap between the two electrodes in order to increase converted power, risks of sparks increase. Therefore, the potential *V* on the electret must be limited because of the breakdown voltage of the air. The chart in figure 12 gives the breakdown voltage ($V_{max}$) of the air as a function of the air gap between two electrodes due to Paschen's law (11). Some experiments [28] have proven that Paschen's law is valid for ambient air until 3µm, and below this distance, the electric field is limited by field emission.

$$V_{max} = \frac{\alpha(pg)}{\ln(pg)+\beta} \qquad (11)$$

With *p* the pressure in atm, *g* the gap distance in meters and $\alpha$ and $\beta$ two constants depending on the composition of the gas. In our case (ambient air $O_2$(20%), $N_2$(80%)), *p*=1atm, $\alpha$=43.6E6 V/(atm.m) and $\beta$=12.8 .[28, 29]

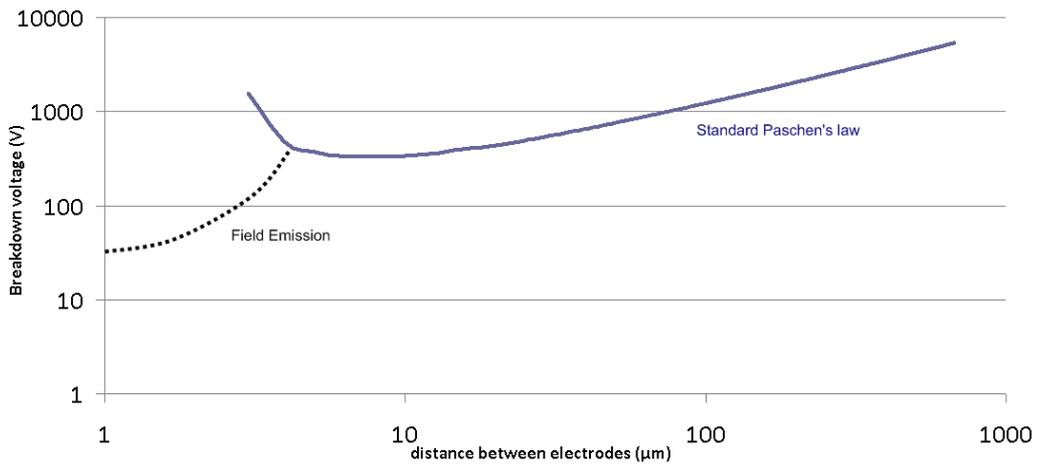

**Figure 12.** Breakdown voltage of the air in function of the distance between electrodes: modified Paschen's law taking field emission into account

For technical reasons (clean-room processes), in the next parts, $g$ will be limited to 5µm. For these dimensions, Paschen's law is valid.

*4.5 Optimisation of the geometrical parameters*

Figure 13(a) presents results of our model for the output power using various values of the bump and the space widths ($b$ and $e$ respectively) while the air gap $g$ and the electret thickness $d$ are kept constant (Each time, the value of $R$ is optimised). Figure 13(b) presents the equivalent results with the simple analytic model (3).

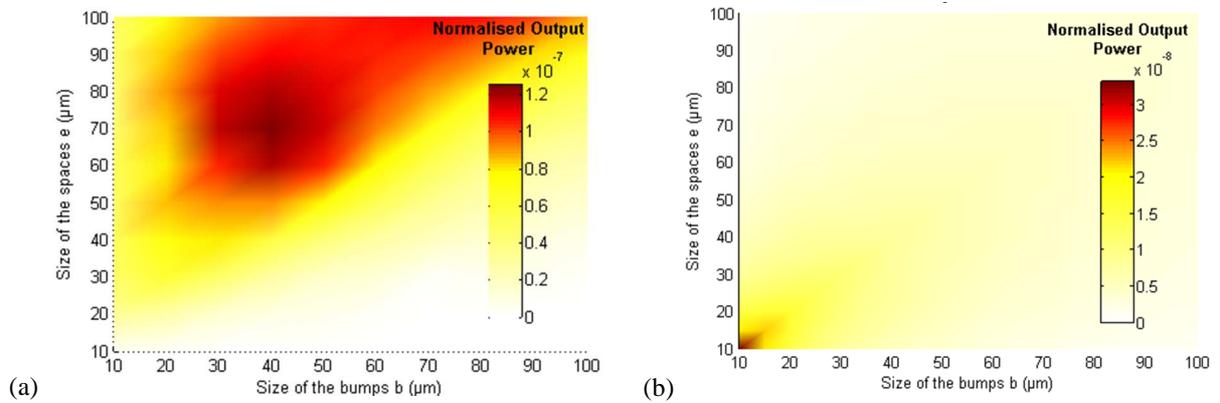

**Figure 13.** (a) Normalised Output Power converted as a function of $b$ and $e$ (FEM) ($g$=5µm, $d$=1µm) and (b) comparison to Boland's formulas

The same protocol is applied to get the output power as a function of ($g$, $d$) while ($e$, $b$) is kept constant. Figure 14(a) is obtained with our model and figure 14(b) is got thanks to (3) (Boland's model).

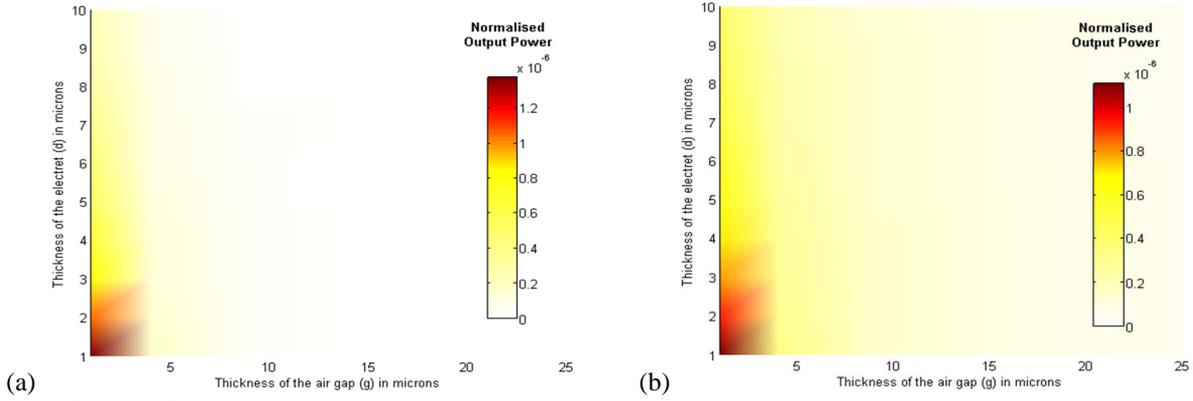

**Figure 14.** (a) Normalised Output Power converted as a function of *g and d* (FEM) (*e*=72µm, *b*=38µm) and (b) comparison to Boland's formulas

FEM model and Boland's model give the same results for the Normalised Output Power as a function of (*g*,*d*) while (*e*,*b*) is kept constant: the output power increases with the decrease of the thickness of the electret and of the air gap. However, for the converted power as a function of (*e*, *b*), the results are very different. Actually, Boland's model states that *dA/dt* has to be maximised to reach the maximum output power. Looking for a maximum area variation per displacement unit induces to choose *e* and *b* as little as possible. By considering fringe effects, it is obvious that it does not maximise the capacitance variation and the output power. A maximum clearly appears for (*e*, *b*); it will be computed using an algorithm of optimisation.

*4.6 Algorithm of optimisation and best result obtained*

The goal of the optimisation is to find the best values for (*e*, *b*, *R*) for a given (*g*, *d*). The following algorithm is applied. Two optimisations are overlapped: the optimisation of *R* is included into the optimisation of (*e*, *b*). These two optimisations use *fminsearch* Matlab function.

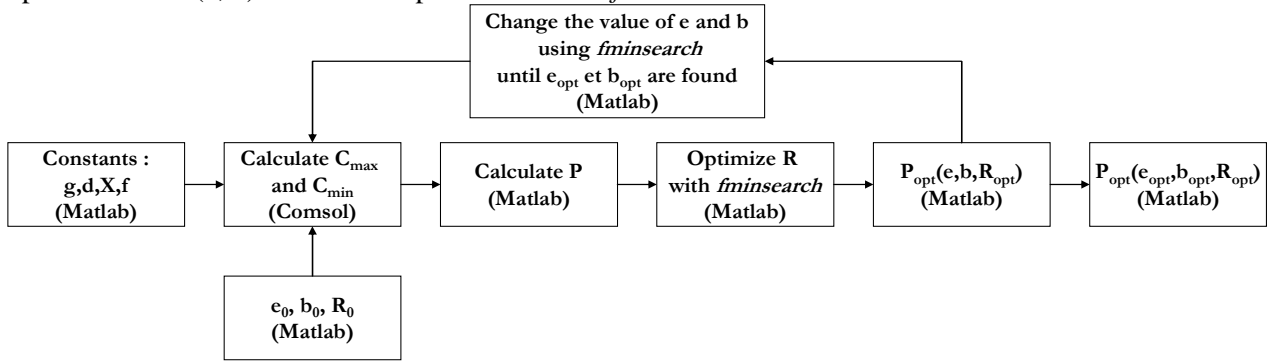

**Figure 15.** Principle of the optimisation for the electrostatic converter

The optimisation starts with $e_0$, $b_0$ and $R_0$, initial values of *e*, *b* and *R*. The value of *g*, *d*, *X* and *f* are considered as fixed parameters of the system and are kept constant during the optimisation. All these parameters are entered in Matlab, which drives the optimisation. In a first loop, Matlab asks Comsol for $C_{min}$ and $C_{max}$ and then computes the converted power and optimises the value of the load to maximise the power output using *fminsearch* function. This loop is inserted into the optimisation of *e* and *b* which is also carried out by the *fminsearch* function. We have chosen to separate the optimisation of *e* and *b* because they are the parameters that can be adjusted the most easily during the fabrication, contrary to *g* and *d* that are given by technological constraints (Figure 15). This separation into two loops limits the complexity of the algorithm and thus the time needed for the optimisation.

Figure 16 presents the geometry that maximises the output power. The gap between electrodes is chosen equal to 5µm for technical reasons (clean room processes and Paschen's effect).

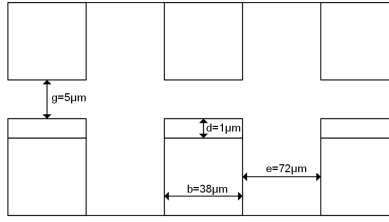

**Figure 16.** Optimal geometric parameters for $(g,d)=(5\mu m, 1\mu m)$

We have optimised the first part of the energy harvester: the geometry of the electrostatic converter and we have also proven that this optimisation is independent of the surface of the converter, the surface voltage of the electret and above all the displacement of the upper electrodes ($x(t)$). The whole electromechanical system can now be optimised.

## 5. Optimisation of the whole electromechanical system

The objectives of this section are to optimise the whole system by using the optimised geometry sized in section 4. Now, the optimisation is not made with the relative displacement ($x(t)$) but with the ambient vibrations ($y(t)$) that create the relative displacement ($x(t)$). Therefore, in this section, $x(t)$ is the response of the mechanical system to ambient vibrations.

### 5.1 Ordinary Differential Equations (ODEs)

The system presented in figure 4 can be modelled by (12). The mechanical equation is deduced from Newton's second law:

$$\begin{cases} \dfrac{dQ_2}{dt} = \dfrac{V}{R} - \dfrac{Q_2}{R} \times \left[\dfrac{1}{C}\right]_{simulation} & \text{(electrostatics)} \\ m\ddot{x} + b_m \dot{x} + kx + f_{elec} = -m\ddot{y} & \text{(mechanics)} \end{cases} \quad \text{with } f_{elec} = \dfrac{d}{dx}\left(\dfrac{Q_2^{\,2}}{2C}\right) \quad (12)$$

To finish the optimisation of the system, we have to optimise both the electrostatic damping and the mechanical structure. The surface of the active part is fixed to 1cm², enabling to get a mobile mass of 1g on a silicon structure. As for the mechanical parameters, we suppose that the vibration is well known and equals to: $y(t) = Y.\sin(2\pi f.t)$, the mobile mass $m$ and the quality factor of the system $Q_{mec}$ are given, and the natural frequency of the device ($f_0$) is tuned to the frequency of the ambient vibrations ($f$). As $b_m = m\omega_0/Q_{mec}$, the global optimisation can be restrained to find $V$ and $R$ that maximise the harvested power.

### 5.2 Effect of V and R on the device

Figure 17 shows the output power as a function of V and R and establishes the existence of an optimum that maximises the harvested power.

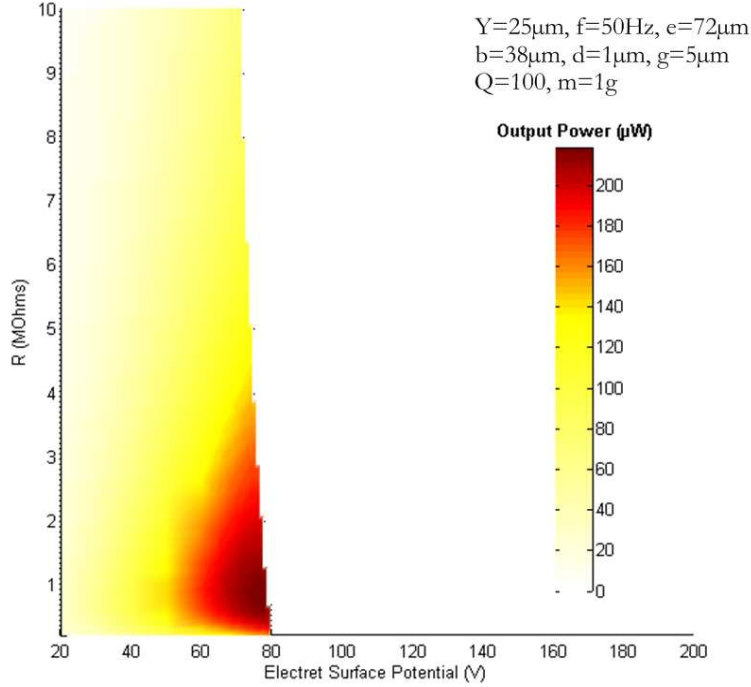
**Figure 17.** Effect of (V,R) on the whole system

Therefore, figure 17 proves the importance of the optimisation of (*V*,*R*) for the entire electromechanical system and shows that, even with ambient vibrations (50µm$_{pp}$@50Hz), up to 200µW can be harvested with an electrostatic energy scavenger using electrets if the quality factor of the mechanical system is good (Q=100). It also verifies that when *V* is too big, it blocks the mobile mass and the harvested power decreases to nearly 0. Finally, the optimisation shows that the maximum harvested power is obtained when *V* is just below the critical limit that causes a blockage of the mass.

*5.3 Comparison to the reference model developed by William and Yates [30] with optimised parameters*

The model developed by William and Yates [30] is the reference model for energy harvesting from vibrations using a resonant system. It considers that energy harvesters can be modelled by a mass (*m*) maintained by springs in a support and damped by a mechanical and an electrical forces, like us. The difference with our model lies on the electrostatic force that is treated as a viscous force. William and Yates have proven that the maximum power that can be extracted from this kind of energy harvester is:

$$P_{max} = \frac{mA^2 \xi_e}{4(\xi_e + \xi_m)^2 \omega_n} \quad (13)$$

where *A* is the acceleration of the ambient vibrations, $\omega_n$ the natural angular frequency of the system, $\xi_m$ the mechanical damping coefficient and $\xi_e$ the electrical damping coefficient. The value of $P_{max}$ is maximum when $\xi_e = \xi_m = \xi = 1/2Q_{mec}$.

Table 2 gives the output power with the model developed by William and Yates and by our model for different masses and different mechanical factors of quality.

**Table 2.** Comparison between our model and William and Yates' model [30]

| | | Our model | | | William and Yates [30] |
|---|---|---|---|---|---|
| m (g) | $Q_{mec}$ | Optimised V (V) | Optimised R (MΩ) | Output power (mW) | Output power (mW) |
| 1 | 10 | 56 | 4.5 | 0.029 | 0.024 |
| 1 | 50 | 75 | 1.3 | 0.142 | 0.121 |
| 1 | 100 | 77 | 0.8 | 0.218 | 0.242 |

| 2 | 10 | 79 | 5.0 | 0.059 | 0.048 |
| 2 | 50 | 106 | 1.2 | 0.284 | 0.242 |
| 2 | 100 | 109 | 0.8 | 0.436 | 0.484 |
| 5 | 10 | 126 | 5.0 | 0.147 | 0.121 |
| 5 | 50 | 168 | 1.2 | 0.711 | 0.606 |
| 5 | 100 | 172 | 0.8 | 1.091 | 1.211 |
| 10 | 10 | 178 | 4.9 | 0.294 | 0.242 |
| 10 | 50 | 239 | 1.2 | 1.421 | 1.211 |
| 10 | 100 | 242 | 0.8 | 2.200 | 2.422 |

Table 2 shows a good correlation between our result and the results given by the model developed by William and Yates. The differences observed come from the fact that the electrostatic force in electret-based energy scavengers cannot be simply modelled by a viscous force. It proves moreover that, with optimised ($V$, $R$) the output power is proportional to the mobile mass. Therefore, with ambient vibrations (50µm$_{pp}$@50Hz), it seems possible to get up to 218µW with an active surface of 1 cm² and a mass of 1g. This corresponds to a figure of merit of 112, which is 8 times better than the best structure in the state of the art. It also corresponds to a power density of 218µW/gram with a quality factor of 100 and 29µW/gram for a quality factor of 10. Table 2 also proves that for MEMS ($m \approx 1$g), the use of high voltage electrets is not useful (100V are largely sufficient), as soon as the gap is small ($g \approx 5$µm). This value can be for example obtained on simple $SiO_2$ corona electrets that can keep up to 10mC/m² (350V/µm) [31].

Finally, we have proven the validity of William and Yates' model even for electrostatic energy scavengers using electrets. Indeed, even if the electrostatic force in our converter cannot be simply modelled as a viscous force, the results in terms of output power are similar to the output power in William and Yates' model. Therefore, William and Yates' model is a good approximation for estimating the available power whatever the technology used (piezoelectric, electrostatic, electromagnetic), as soon as the system is resonant (i.e. the energy absorbed in each cycle is small compared to the mechanical energy stored in the mass-spring system).

## 6. Conclusion and Perspectives

We modelled an in-plane multibumps electret-based energy scavenger with a FEM software. We proved that the variable capacitance can be simply expressed with a cosine function and the determination of two extreme values: this greatly limits computation times compared to point-by-point simulation. It has also been proven that the electrostatic part of the system and especially the size of bumps and spaces should be optimised to maximise the output power of the energy harvester. The optimised structure established that, up to 218µW could be obtained with ambient vibrations: this gives the best result with our factor of merit. The optimisation finally proves that electrets that can keep high surface voltage on small thicknesses are particularly well adapted for energy harvesters: $SiO_2$-based electrets meet this criterion.

To conclude, even if electrostatic energy harvesters are at a less advanced stage of development compared to electromagnetic or piezoelectric systems, this study proves that they are as interesting as the other solution quoted above in terms of harvested power as long as the gap can be in the order of magnitude of some microns.